# Role of Wadsley Defects and Cation Disorder to Enhance MoNb$_{12}$O$_{33}$ Diffusion

CJ Sturgill[1†], Manish Kumar[1†], Nima Karimitari[1†], Iva Milisavljevic[2†], Coby S. Collins[1], Aaron Hegler[1], Hsin-Yun Joy Chao[3], Santosh Kiran Balijepalli[4], Scott Misture[2*], Christopher Sutton[1*], Morgan Stefik[1*]

† C.S., M.K., N.K., and I.M. contributed equally to this paper.
   1) Department of Chemistry and Biochemistry, University of South Carolina, Columbia, SC 29208, USA.
   2) Inamori School of Engineering, Alfred University, Alfred, NY 14802, USA.
   3) Center for Nanophase Materials Sciences, Oak Ridge National Laboratory, Oak Ridge, Tennessee 37831, USA.
   4) Office of the Vice President for Research, University of South Carolina, Columbia, SC 29208, USA.
   *Corresponding Authors: misture@alfred.edu, CS113@mailbox.sc.edu, morgan@stefikgroup.com

**Abstract**
Wadsley-Roth (WR) niobates have emerged as high-rate anode materials that can combine rapid ionic diffusion with good electronic conductivity. WR compounds have been defect-enhanced by limited annealing, however, such materials often contain multiple types of defects. In particular, both Wadsley defects (variable block size) and transition metal disorder have the potential to modify transport rates, however the corresponding effects are not well understood mechanistically. Here, MoNb$_{12}$O$_{33}$ (MNO) was calcined at two different temperatures to compare a defect-rich condition (MNO-800) with a proximal order-rich condition (MNO-900) as assessed through XRD, XANES, EXAFS, and STEM characterizations. Galvanostatically cycled lithium half cells of MNO-800 exhibited additional capacity (307 mAh/g at 0.1C, 4.66% higher) and improved high-rate capacity of 200 mAhg$^{-1}$ at 10C. ICI-based overpotential analysis identified solid state diffusion as the dominant rate limiting process where MNO-800 correspondingly exhibited ~3X faster capacity-weighted diffusivity. A machine-learning interatomic potential was trained to density functional theory and then applied with molecular dynamics (MLIP-MD) to examine the possible roles of Wadsley defects and transition metal disorder. For both defect-types, Li was found to populate and activate fast diffusion paths from window sites at lower extents of lithiation as compared to the order-rich model.

**Introduction**
With the ever-growing need for durable high-power batteries, WR niobates are an excellent class of anode materials that often have rapid transport characteristics, suitable working potentials for long term electrolyte stability, and often exhibit second-order phase changes upon lithiation to limit cracking during repeated cycling.[1,2] WR materials are crystallographic shear structures comprised of corner-sharing ReO$_3$-like octahedra arranged in blocks surrounded by edge-sharing shear planes.[1,3] Such phases with a single block size may be E-type with edge-sharing only, T-type with a tetrahedrally coordinated metal in the block corners, or M-type for a mixture thereof.[3,4] Ionic diffusion generally has the lowest activation energy along the block structure, in part due to the multiplicity of parallel channels enabling redundant paths around defects.[5–7] Additionally, these phases generally become metallic upon early stages of lithiation.[2,5,8] The structure changes upon lithiation include expansion along certain crystallographic directions that can be partially compensated by contraction along other directions to limit the overall strain and preserve the structure through repeated cycling.[2,9] $MoNb_{12}O_{33}$ ("MNO") recently emerged as a high-performing T[3×4] WR niobate with a reversible 200 mAhg$^{-1}$ capacity at 10C.[10] Recently the first single crystal measurements of MNO were reported where the corresponding powder exhibited a reversible capacity to 253.3 mAhg$^{-1}$ at 0.1C. Notably that study included diffusion analysis compared to the isostructural tantalate analog.[6]

Defect engineering has been used to enhance ionic and electronic transport in battery materials.[11–15] More specifically for WR materials, Wadsley defects are intergrowths that form as extra/missing slabs of octahedra within the block structure (i.e. block size change) and their presence can vary by synthesis strategies.[16,17] These intergrowths are suggested to form as a means of initial defect elimination, strain relief, and accommodation of non-stoichiometry.[18–20] Though Wadsley defects are often eliminated with further annealing, they can also entropically stabilize the structure.[21] Previous studies on M[3×4, 3×5] H-$Nb_2O_5$ noted that a sample rich with Wadsley defects (950 °C) exhibited less volume expansion, increased capacity, and faster ionic diffusion as compared to its pristine counterpart (1,300 °C).[22,23] This is consistent with the notion of increased block size trending with increased diffusivity as intergrowths usually lead to some blocks with increased sizes.[24] Multi-metal WR compounds are known to alter the site-preference of different metals as a function of the block size which convolves block size changes with cation ordering changes.[9,24–28] There is a prevailing design principle that cation disorder will suppress lithium ordering and enable accelerated lithium diffusion.[9,29–32] Cation ordering, however, is both a function of the equilibrium structure for a given composition in addition to its kinetically-limited organization during synthesis. Thus, the study of multi-metal WR compounds with different calcination conditions are likely to simultaneously vary the Wadsley defects and cation ordering. A recent study of a defect-rich titanium niobium tungsten oxide identified Wadsley defects and variable presence of tetrahedra that led to remarkable performance.[25] Cation disorder is likely in that report given the non-equilibrium structure, however, the site occupancies were not well supported empirically and the effect of cation disorder was not addressed computationally. Another similar study on a disordered $Nb_{12}WO_{33}$ concluded that there were wider diffusion pathways and changes in local structure that improved lithium diffusion.[33] Therefore, further understanding the origins of such structure-property relationships will require parsing the effects for these concomitant changes.

In this work, we aim to understand the separate effects of Wadsley defects and cation disorder for WR materials using $MoNb_{12}O_{33}$ as an example. A defect-rich condition (MNO-800) and an order-rich condition (MNO-900) were prepared at different temperatures and were structurally characterized using XRD, Rietveld refinement, XAS (EXAFS/XANES), and STEM. The corresponding electrochemical properties, rate limiting step analysis, and lithium diffusivities were measured and compared to three types of computational models to separate understand the behavior of a defect-free structure as compared to models with Wadsley defects and models with varied cation placement alone.

## Results

*Structural Characterization of MoNb$_{12}$O$_{33}$ Samples*

A sol-gel strategy was used to prepare MNO using two calcination temperatures (800 and 900 °C) to vary the extent of crystallization and the quantity of defects. Here the utilized sol-gel strategy has the advantage of starting with molecular-level mixing to minimize the amount of diffusion needed for crystallization. Both samples were prepared from the same sol-gel batch, thus the heat treatment was the only synthetic variable. After synthesis, the resulting MNO-800 and MNO-900 were characterized in detail. The powder XRD analysis followed by Rietveld refinement indicated that both samples were single-phase materials that correspond to the monoclinic MoNb$_{12}$O$_{33}$ crystal phase (Figure 1a,b). The refinement results suggested increased crystallite microstrain in MNO-800 that was almost two times higher than in MNO-900 (Table S1), which demonstrates some large content of point and/or extended defects in MNO-800. A detailed analysis of the Nb K-edge and Mo K-edge XANES spectra of the samples before and after lithiation, including pre-edge peak fitting and comparison with measured standards, is presented in Figure S1. The Nb K edge (Figure 1c) EXAFS oscillations indicate very minor differences between the samples in the local structure around the Nb absorber atom in both first (R=1−2.5 Å) and the second coordination shells (R=3−4 Å). Conversely, the Mo K edge (Figure 1d) EXAFS show the presence of only the low R (R<2.7 Å) scattering paths from the first coordination shell oxygen atoms around the Mo absorber, with the intensity of the peak reducing with annealing temperature. Fitting of the Mo K EXAFS data using the theoretical crystalline MoNb$_{12}$O$_{33}$ structure (Figure S2) was successfully performed for MNO-900. The fitting results from EXAFS showing the variance in the coordination number (N), half-path length (ΔR), and the mean square displacement ($\sigma^2$) (Table S2), revealed a significant increase in the first-shell Mo–O bond length, from ~1.70 Å in the model structure to ~1.78 Å in MNO-900. The average coordination number was ~4.2, indicating predominantly four-fold coordination of Mo. In contrast, attempts to fit the MNO-800 EXAFS data using existing literature or theoretical models were unsuccessful, suggesting that the Mo first shell coordination is not uniformly tetrahedral or octahedral and that some more complex model is needed, perhaps one that includes extended crystalline defects. A direct comparison of the Mo K edge EXAFS FT in R-space shows a markedly higher first-shell peak intensity for MNO-800, suggesting a substantially higher average coordination number than in MNO-900, where the coordination is reasonably described as tetrahedral. The MNO-800 dataset also exhibits peaks in the 2–3 Å range, consistent with strong scattering contributions from neighboring atoms at intermediate distances. Together, these observations indicate that the local Mo environment in MNO-800 is highly defective due to incomplete crystallization at the lower annealing temperature, and that Mo is not necessarily restricted to four-fold coordination.

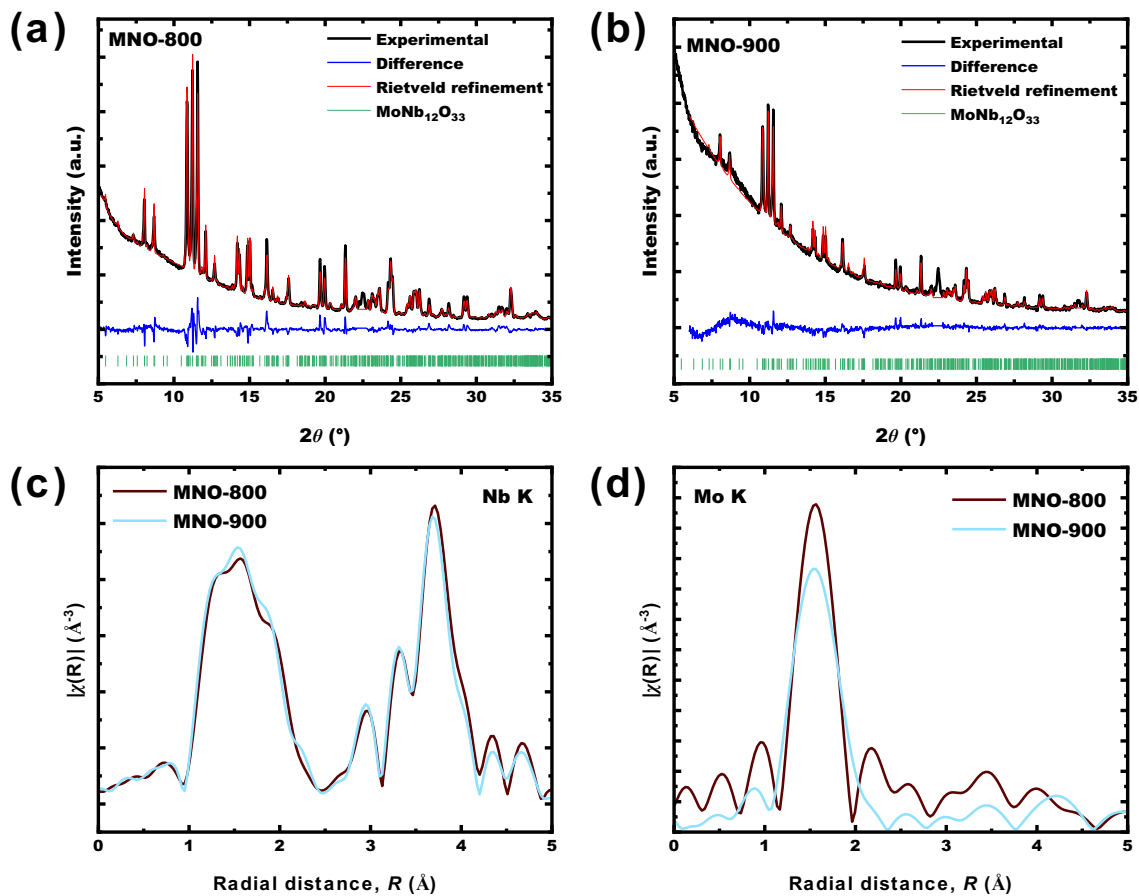

Figure 1. XRD and Rietveld refinement results for (a) MNO-800 and (b) MNO-900 samples (excluded sample holder peak at ~22.5). (c) Nb K edge and (d) Mo K edge radial distribution function of FT EXAFS oscillations, all phase-corrected.

Kinetically limited crystallization is well known to result in transient (non-equilibrium) defect-rich states. The early 1970's studies of WR materials often relied on electron microscopy to directly observe the atomic structure since X-ray diffraction often misses Wadsley defects and WR structures are furthermore infinitely adaptive to composition changes.[34,35] HAADF-STEM measurements were used to observe the crystal structures directly in real space. Samples were generally oriented with the (010) plane perpendicular to the beam path to image the block structure face-on. The STEM data for MNO-800 (Figure 2a-f) included observations of 9 different block sizes and a variety of interface types. Notably most electron micrographs of MNO-800 included defects within each field of view. Figure 2a shows a highly defective region with block sizes ranging from [2×5] to [4×6]. Interestingly, broken shear planes were observed where some adjacent blocks were interconnected. A doubled shear plane was also observed that was 3 atoms wide (Figure 2a). The next figure panels (Figure 2c,e) show the expected [3×4] block sizes in addition to [3×5], [4×4], and [5×5]. Notably these block size defects also have mixed presence of tetrahedra with both T-type and E-type interfaces. In contrast, the data from sample MNO-900 (Figure 2g-l) predominantly exhibited T[3×4] blocks (Figure 2g,k). A minor amount of [3×5] blocks were found with a mixture of E-type and T-type interfaces (Figure 2i). Additional wide field of view STEM images (Figure S3 and Figure S4) were consistent with these trends for defect-rich MNO-800 and order-rich MNO-900.

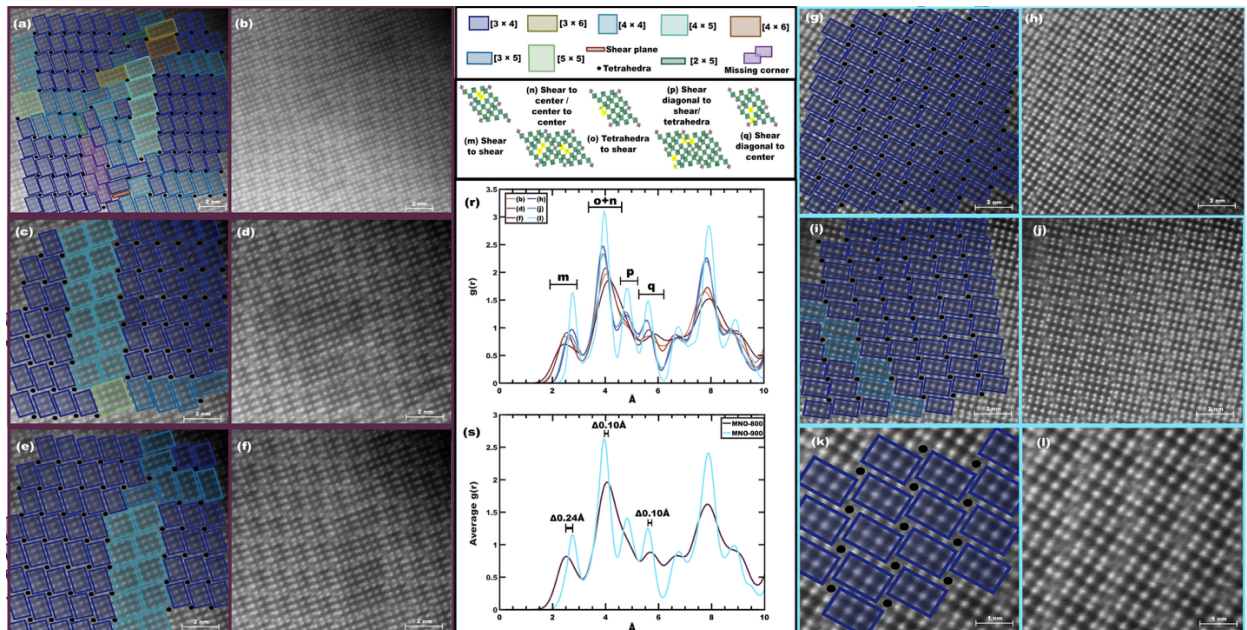

Figure 2. HAADF STEM data for MNO-800(a-f, left side) and MNO-900(g-l, right side). Each region is presented with block overlays (a,c,e,g,i,k) indicating the blocks colored according to size and without overlays (b,d,f,h,j,l). Specific bonds (m-q) were associated with the radial distribution functions (r,s). Single-image RDFs (r) from calculated from the indicated insets. Average RDFs for each sample type (s) were compared. The RDF analysis (r,s) corresponds to the 2D metal distribution in the (010) plane.

The local structure and atomic spacings were analyzed in further detail. Each atom coordinate (see Figure S5 for coordinate overlay) was identified for the images in Figure 2 for radial distribution function (RDF) calculations. Please note that these STEM based RDFs correspond to the 2D sample projections rather than 3D atomic spacings as measured by EXAFS. The RDFs of these six STEM images (Figure 2r) are similar for each condition and were averaged (Figure 2s) to improve statistics. Four regions under 6 Å were attributed to specific atom pairs (Figure 2m-q). The first RDF peak for MNO-800 corresponded to 0.24 Å shorter separation of edge-sharing octahedra. The second RDF peak for MNO-800 corresponded to 0.10 Å longer distances on average between corner-sharing octahedra, including octahedra within the block as well as those connected to tetrahedra. This and the practically unchanged lattice constants from Rietveld refinement (<0.005 Å differences) are consistent with a balanced combination of localized contractions with localized expansions. Notably, the expansion between corner-sharing octahedra provides additional room for lithium hopping between window sites (Figure 2q). Additionally, the broadening/disappearance of peak-p from 4.4-5.2 Å for MNO-800 (Figure 2s) is consistent with the heterogeneity of mixed E-type and T-type block interfaces (Figure S6).[3] The collection of data thus demonstrates significant structural differences between the defect-rich MNO-800 and the order-rich MNO-900.

Several defect-rich WR syntheses have been reported that include mixed phases, mixed interface types, block size variation, and multiple transition metal cations. Some of these prior reports have claimed enhanced diffusion that was attributed to 1) the presence of planar defects,[36] 2) on average larger block sizes with the presence of T-type and E-type interfaces[25], or 3) similar shear-plane contraction paired with block expansion[33]. The literature to date has not yet recognized how these defects necessarily lead to additional point defects that obfuscate clear evaluations of cause-and-effect. For example, the increase of block size under constant composition necessarily leads to cation disorder as well as oxygen vacancies, interface change, or localized inhomogeneity to preserve charge balance. Here MNO-800 and MNO-900 were validated to have the same nominal composition of $MoNb_{12}O_{33}$ by EDS (Figure S7 and Table S3) that would place one Mo atom in each T[3×4]. Additionally, XPS confirmed that both samples had fully-oxidized $Mo^{6+}$ and $Nb^{5+}$ (Figure S8). When this composition (1 Mo per 12 Nb) is placed within larger block sizes (e.g. T[5×5] with 26 atoms per block) there will be more than one Mo atom within each unit cell on average. The Mo content of each unit cell thus exceeds the number of tetrahedral positions and leads to Mo placement within a corner or edge sharing octahedra of the block. For example, STEM-EDS mapping (Figure S9) of MNO-800 identified some Mo presence throughout [3×8] blocks with enrichment along edges and shear planes. Furthermore the 69 oxygen positions with this example T[5×5] structure cannot be fully occupied (oxygen vacancy) when constrained by the composition (1 Mo per 12 Nb) and the oxidation states ($Nb^{5+}$, $Mo^{6+}$). This mismatch curiously requires two comparisons to account for the separate effects of block size and cation ordering changes. Thus, there remains ambiguity as to how each of these defects alters lithium diffusion.

*Electrochemical Properties of $MoNb_{12}O_{33}$ Conditions*
Lithium half-cells were assembled to compare the electrochemical properties of both samples. Galvanostatic cycling was performed at variable C-rates between 1.0-3.0 V vs $Li/Li^+$ to test the rate-dependent capacities. Constant current was followed by a voltage hold after both lithiation and delithiation to ensure end point equilibration. The voltage profiles are shown for the first four cycles (Figure 3a,b) for each condition. Both samples exhibited irreversible charge transfer during the first cycle that was consistent with lithium trapping. The irreversible current from the first cycle was consistent with x=1.6 and x=1.4 lithium atoms ($Li_xMoNb_{12}O_{33}$) trapped in MNO-800 and MNO-900, respectively. The reversible capacity for MNO-800 was x=19.9 at 0.1C, corresponding to 1.59 Li/metal atom. By comparison, the reversible capacity for MNO-900 was x=19.1 at 0.1C, corresponding to 1.52 Li/metal atom. The differential capacities of the first and second cycles were compared (Figure 3c,d) where both conditions exhibited irreversible changes, most notably during lithiation in the ~1.90-2.20 V vs $Li/Li^+$ region. Previous studies with $MoNb_{12}O_{33}$ have attributed dQ/dV peaks from 1.10-1.30V to $Nb^{4+/3+}$ redox, peaks from 1.50-1.80V to a convolution of the $Nb^{5+/4+}$ and $Mo^{5+/4+}$ redox, and finally peaks from 1.90-2.20V to the $Mo^{6+/5+}$ redox.[6,10] Comparing the differential capacities of both conditions reveals that MNO-800 has additional charge storage from ~1.15-1.30 V and also from ~1.90-2.10 V, consistent with further reduction of Nb/Mo (Figure S10a). Notably a similar capacity increase was noted elsewhere for WR $Nb_2O_5$ with Wadsley defects.[23] Lithiation capacities were examined at different current densities (Figure 3e). The formation cycle at 0.1C exhibited the highest lithiation capacity, interpreted as the initial trapping of lithium, similar to some other WR phases.[10,37] Subsequent cycling at a rate of 0.2C led to a reversible lithiation capacity of 280 and 272 $mAhg^{-1}$ for MNO-800 and MNO-900, respectively. Cycling at progressively higher C-rates led to polarization that gradually reduced the capacities during the galvanostatic phase. Generally, MNO-800 exhibited improved capacity retention at high rates with a remarkable 226 mAhg$^{-1}$ at 5C that significantly exceeded that for MNO-900 (182 mAhg$^{-1}$). Returning to cycling at 0.2C, both conditions were within the error of their initial 0.2C lithiation capacities. The coulombic efficiencies remained >96% (N≥3) when including the charge passed during the voltage hold periods (Figure S10b). MNO-800 was further cycled with progressively higher C-rates (Figure 3f) with remarkable lithiation capacities of 200 and 151 mAhg$^{-1}$ at 10C and 20C, respectively, both under 10 min. This compares well to a report of presumably defect-rich microspheres of MoNb$_{12}$O$_{33}$ that exhibited a capacity of 200 mAhg$^{-1}$ at 10C.[10] Notably, the MNO-800 performance reported here also compares favorably with leading benchmarks such as T[4×5] Nb$_{16}$W$_5$O$_{55}$, which had 128 mAhg$^{-1}$ capacity at 20C.[2] A Ragone plot enabled comparison of the power density and energy density of MNO-800 to other anode materials including graphite, Li$_4$Ti$_5$O$_{12}$, and several benchmark WR niobates (Figure 3g). The calculations, with specifications detailed in Table S4, presumed a standard 4V cathode and importantly utilized capacity-voltage curves for each C-rate to realistically account for anode polarization. The WR niobates TiNb$_2$O$_7$, Nb$_{16}$W$_5$O$_{55}$, and Nb$_{18}$W$_8$O$_{69}$ were largely overlapping in this analysis with MNO-800 surpassing each. Additional long-term cycling of MNO-800 at 20C (3-min (dis)charge) resulted in ~93%, ~82%, and ~74% capacity retention through 1k, 5k, and 10k cycles (Figure 3h), respectively. This longevity indicates these materials are impressively stable despite the strain gradients from rapid and high-power cycling.[38] Operando WAXS of both samples lithiated to x=13 (1 Li/TM) revealed 4.9 vol% expansion for MNO-800 and 7.8 vol% for MNO-900 (Figure S11 and Table S5), consistent with prior reports of defect-rich WR materials having less volume expansion upon lithiation.[23,25,33]

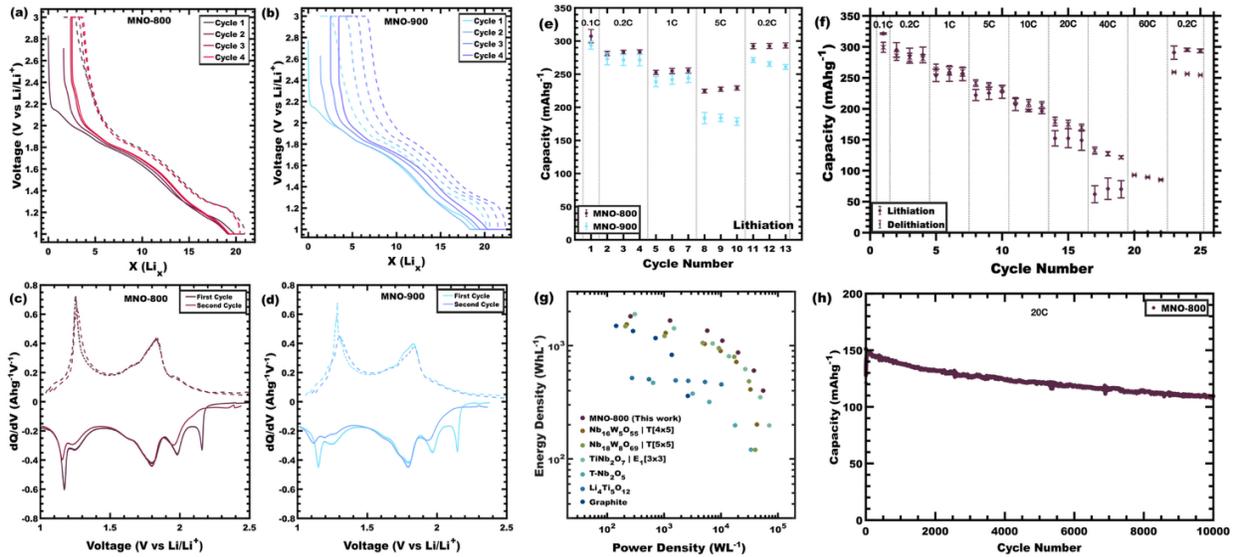

Figure 3. Galvanostatic voltage-capacity profiles of the first 4 cycles of (a) MNO-800 and (b) MNO-900 (0.1C then 0.2C three times). The differential capacity profiles (c,d) for the first and second cycles are shown. Lithiation (solid lines) and delithiation (dashed lines) data are shown. Galvanostatic lithiation (e) capacities were measured at different C-rates for MNO-800 and MNO-900. MNO-800 was measured at higher current densities (f) and the results were compared to literature precedents on a Ragone plot (g). A standard 4V cathode was assumed and the voltage was taken as the average at each current density. Data sources: Nb$_{16}$W$_5$O$_{55}$ [2], Nb$_{18}$W$_8$O$_{69}$ [39], TiNb$_2$O$_7$ [40], T-Nb$_2$O$_5$ [41], Li$_4$Ti$_5$O$_{12}$ [42], and Graphite [43]. Long-term cycling (h) of MNO-800

at 20C was measured through 10,000 galvanostatic cycles. Statistical data presented as the mean with the error-of-the-mean.

Intermittent current interruption was used to compare overpotentials of possible rate-limiting steps and to quantify ionic diffusivity.[44] As described previously (Figure S12), the voltage relaxation during current interruptions can be parsed into an instantaneous overpotential ($\Delta V_{2ms}$), corresponding to electronic resistances and bulk electrolyte resistance and also the charge-transfer overpotential ($\Delta V_{CT}$).[45] The diffusion overpotential ($\Delta V_D$) was determined similarly by referencing a pseudo-equilibrium profile adjusted for hysteresis (Figure S12).[45,46] Both conditions had dominant diffusion overpotentials that ranged from 50-170 mV (Figure 4a,b). The $\Delta V_{CT}$ values were also similar, ranging from 6-13 mV. In contrast, the $\Delta V_{2ms}$ for MNO-900 was 12-18 mV, whereas that for MNO-800 was several times lowers at 4-7 mV. Given that the cell construction was replicated for both conditions, the change in $\Delta V_{2ms}$ is best explained by changes to the electrical conductivities with MNO-800 having lower resistance. Regardless of this difference, solid state diffusion was by far the dominant rate-limiting process, so diffusivities were next compared. The diffusion length scale was determined using Porod analysis of SAXS data (Figure 4c,d) to determine the specific surface area of each condition.[47,48] MNO-800 and MNO-900 had 1.63 and 1.40 $m^2g^{-1}$ of specific surface area, respectively. The slightly lower surface area of MNO-900 is consistent with some additional coarsening with the higher temperature crystallization.

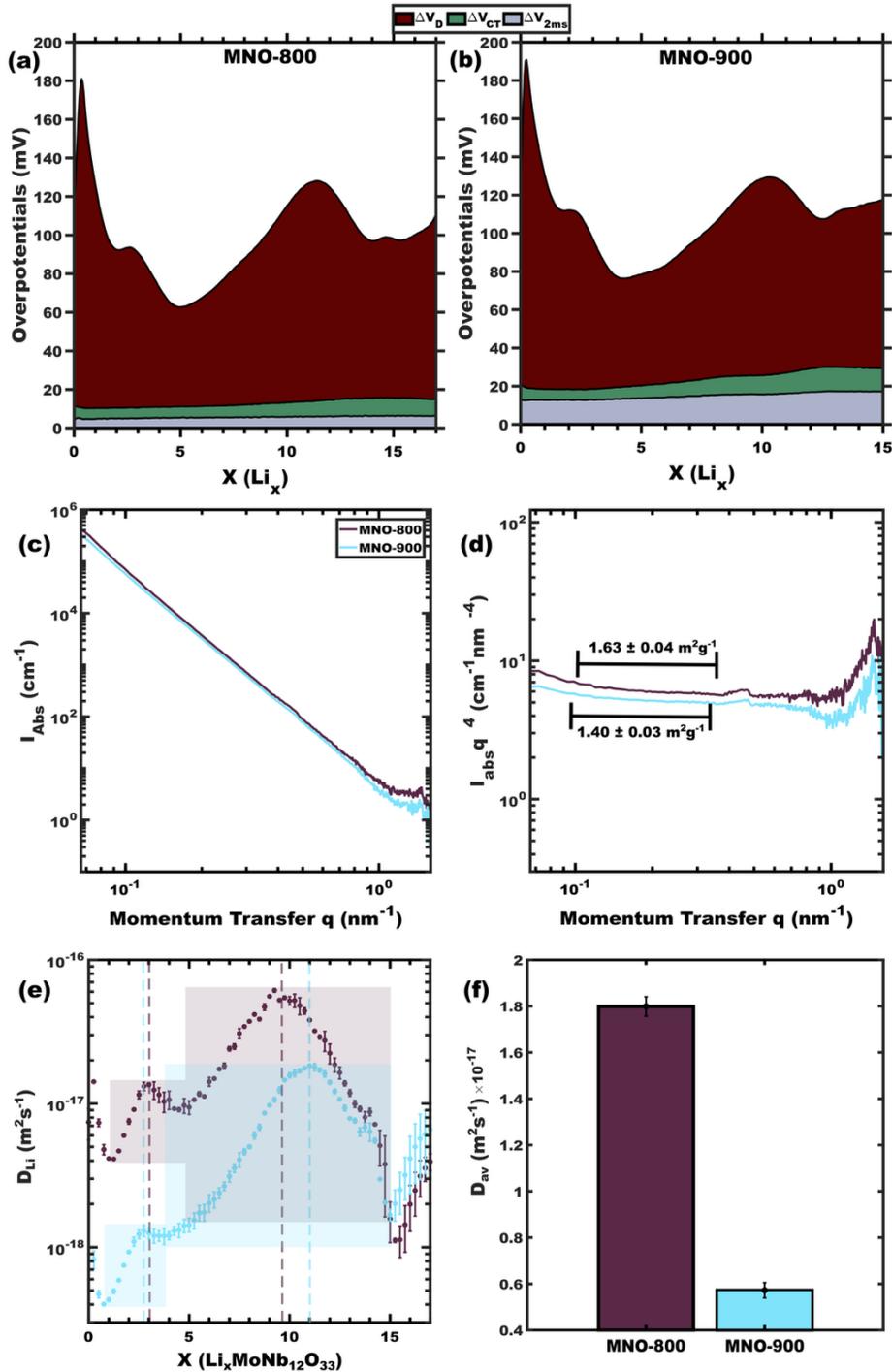

Figure 4. ICI-derived overpotentials for (a) MNO-800 and (b) MNO-900 measured at 0.1C. Absolute intensity small-angle X-ray scattering data (c) of both samples were used for Porod analysis (d) to calculate the specific surface areas. The surface area and ICI analysis were used to calculate lithium diffusivities (e) at each extent of lithiation. Shaded boxes highlight parabolic diffusion profiles to clarify width and position differences. The figure-of-merit termed capacity-weighted average diffusivity (f) was calculated for comparison.

The surface area and ICI analysis were used to calculate lithium diffusion coefficients at each extent of lithiation (Figure 4e). The diffusivities each varied by roughly an order of magnitude with MNO-800 generally having an order of magnitude larger diffusivity until the D(x) values converged around x~11-14. Both materials exhibited multi-parabolic D(x) trends that were concave down, consistent with progressive lithium ordering through different site types. For each site type, diffusivity increases as the site becomes more populated and then decreases as the site becomes crowded, consistent with a Frumkin isotherm and lithium-lithium repulsion.[49,50] A similar set of parabolic D(x) trends were found recently for $VNb_9O_{25}$, however, in that case, the fastest diffusivity was found at the onset of lithiation x~0.[46] In contrast, both MNO conditions exhibited the fastest diffusivity at x~9-11. The parabola locations also provide information about the occupancy of each site type. Both materials have an initial parabola that is Δ(x)~1 in width. The second parabola widths are different, however, with Δ(x)~4 and Δ(x)~3 for MNO-800 and MNO-900, respectively. The third parabola widths scale opposingly with Δ(x)~10 and Δ(x)~11 for MNO-800 and MNO-900, respectively (Figure 4e, shaded regions). These different ratios suggest that MNO-800 and MNO-900 differ from each other in terms of the number of each occupied site type, e.g., one having more occupied shear plane pocket sites and the other having more occupied interior block window sites, *vide infra*. In contrast, near the end of full lithiation the two parabolas located at x~13-15 and then x>15 exhibit very similar D(x) values, consistent with closely related diffusion paths and crowding effects (high site occupancies). The comparison of diffusivities for materials operating over a wide voltage range (e.g., second order phase change materials) is best done with a comprehensive figure-of-merit such as capacity-weighted average diffusivities ($D_{av}$).[51] The capacity-weighted average diffusivities for MNO-800 and MNO-900 were calculated as $1.80×10^{-17}$ and $5.72×10^{-18}$ $m^2s^{-1}$, respectively. This corresponds to 3X higher overall diffusivity for MNO-800 (Figure 4f). MNO-800's performance advantages are thus substantially associated with its defect-enhanced lithium diffusivity during the first stages of lithiation and slightly due to its improved electronic conductivity.

*Computational Analysis and Simulations of $MoNb_{12}O_{33}$ Conditions*
To understand the separate roles of cation disorder and Wadsley defects we considered three types of models: (Type-1) a pristine T[3×4] model (Figure 5a) having the formula $MoNb_{12}O_{33}$ where the Nb are placed at the octahedral sites in the block with Mo at the tetrahedral sites on the corner of the block; (Type-2) several cation disordered T[3×4] structures in which the Mo cation is placed at different positions in a double unit cell. Model T[3×4-dis-1] (Figure 5b and Figure S13) had mixed Mo/Nb placement at the tetrahedral sites with the liberated Mo placed at the octahedral position in the block middle. Model [3×4-dis-2] also had mixed tetrahedral occupation but placed the relocated Mo at the corner of the block. Model [3×4-dis-3] placed Nb at the tetrahedral sites and both Mo at the octahedral positions, with one Mo at the corner and another in the middle of the block T[3×4-dis-3]. Similarly model T[3×4-dis-4] had Nb in tetrahedral sites with both Mo at the middle of the block whereas model T[3×4-dis-5] placed both Mo across the shear plane. The latter corresponds to the configuration state selected based on the special quasirandom structure (SQS) procedure.[52,53] The illustrative figures for the placement of the Mo atoms are provided in Figure S13, and their decomposition enthalpy are provided in Table S6. These structural models provide an understanding of the effect of cation disorder on both diffusion and the activation barriers. (Type-3) Several models with different block sizes were considered. A T[3×5] model ($MoNb_{15}O_{40.5}$) (Figure 5c) was included since this block size was frequently observed experimentally. Furthermore, a T[3×4-3×5] model ($Mo_2Nb_{27}O_{73.5}$) (Figure 5d) was included to simulate diffusion with variable block sizes, as found in MNO-800. Notably the WR phase H-$Nb_2O_5$ also has mixed block sizes of 3×4 and 3×5, with Nb in +5 oxidation state.[54] Larger block sizes naturally have a higher average oxidation number for the metal cations. However, when the block size is expanded under constant metal stoichiometry, there is insufficient cationic charge to support placing oxide anions at all octahedra corners, resulting in what could be considered oxygen vacancies despite the metals remaining fully oxidized. Thus, preserving the Nb/Mo ratio close to 12 with Mo only placed in the tetrahedral positions leads to the models shown in Figure 5c,d.

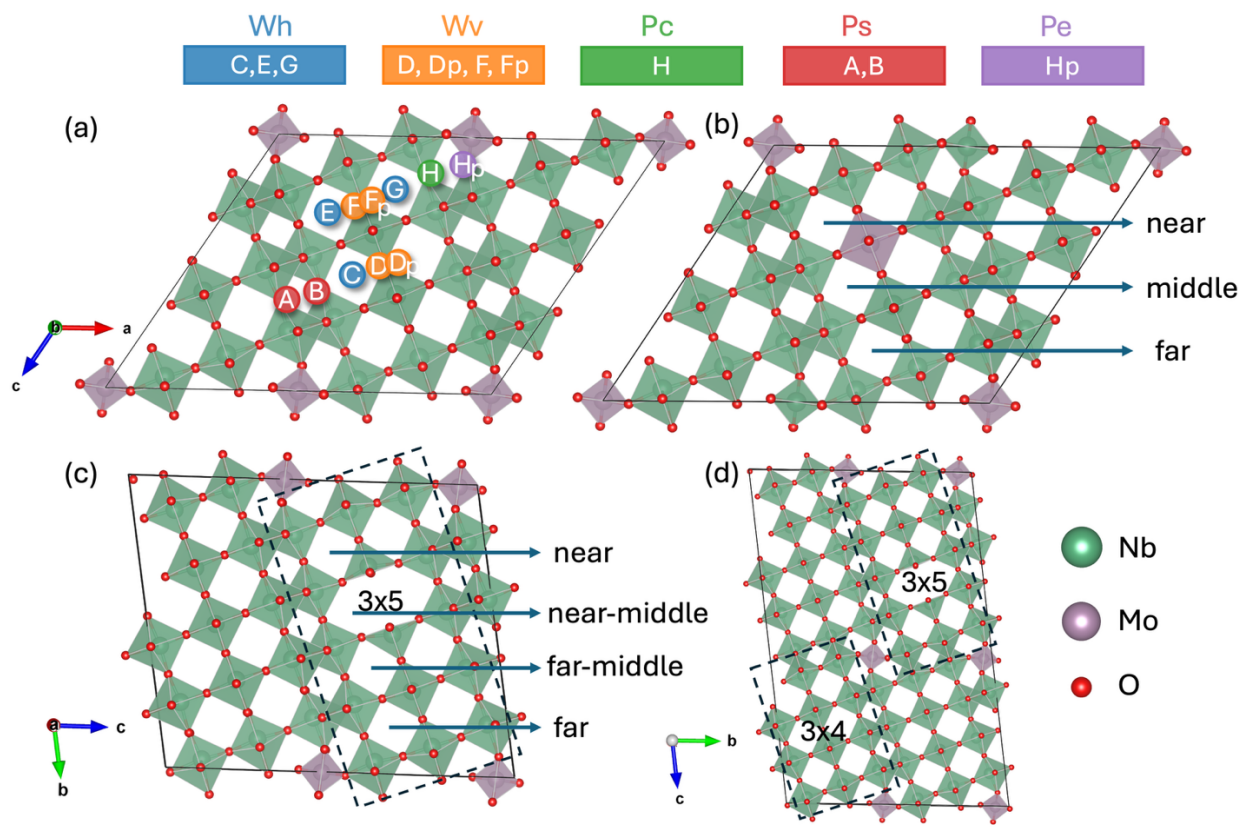

Figure 5. Example model structures analyzed, including T[3×4] (a), T[3×4-dis-1] (b), T[3×5] (c), and T[3×4-3×5] (d). The named Li site-types (Wh-Pe) and their corresponding locations (C-Hp) are indicated. The activation barriers and diffusion from these Li sites were calculated.

The hopping of lithium ions is first considered with the DFT nudged elastic band (DFT-NEB) method to consider hopping within structures having a lone lithium atom. To specify the NEB pathways, we used the nomenclature described by Koçer et al. and as used in our previous work (Figure 5a) for all unique sites.[7,46] A and B are pocket sites at shear plane (Ps); C, E, and G are horizontal windows (Wh); D, Dp, F, and Fp are vertical windows (Wv); H, Hp are pocket sites at corner (Pc) and edge (Pe), respectively. The pocket sites are 5-fold coordinated, and the windows are 4-fold coordinated with oxygen. In comparison to the T[3×4] structure, the T[3×5] and T[3×4-3×5] mixed phases contain similar sites; however, the ratio of these sites changes. Notably, the presence of bigger blocks shifts the ratio of window sites to pocket sites from 0.93 (13 windows, 14 pockets) in the [3×4] block to 1.03 (31 windows, 30 pockets) in the [3×4-3×5] block, and finally to 1.13 (18 windows, 16 pockets) in the [3×5] block structure. This ratio can affect the overall diffusion in the bulk materials because each of these sites has an associated diffusion constant, as described below.

In the pristine T[3×4], the activation barrier ($E_a^{NEB}$) for hopping from horizontal to vertical windows (E↔F) of 67 meV is smaller than the hopping between middle windows (C↔D) at 169 meV, and is noticeably smaller than the pocket site pathway (B↔C) at 510 meV. According to the Arrhenius equation, the lower energy barrier for E↔F corresponds to faster diffusion of Li in this pathway (see Table S7 for all $E_a^{NEB}$). However, given that Li diffusion can be a combination of different pathways, to evaluate the diffusion coefficient and compare with experimental trends, we performed MD simulations at a range of different temperatures and at different starting positions for Li atoms. Given that *Ab initio*-MD (AIMD) simulations are impractical for nanosecond MD simulations, we used a machine-learned interatomic potential (MLIP) that is fine-tuned from MACE-MP0. As detailed in the methods section, for the validation of the MLIP, we performed MLIP-NEB for T[3×4] and T[3×5], and achieved an MAE accuracy of 10.3 and 14.9 meV for all pathways, respectively. As shown in Table S8, this accuracy is enough to capture the same trend in $E_a^{NEB}$ as in DFT-NEB (see Table S8), and therefore suitable for diffusion estimations using MD simulations. In the pristine T[3×4], the diffusion from the windows (E, F) other than the middle of the block is an order of magnitude faster in comparison to the middle windows (C, D), and the diffusion from pocket sites is still slower by about four orders of magnitude (Figure 6a and Table S9). This is consistent with the calculated activation energies via NEB, $E_a^{NEB}$, which indicates that the barriers for these hops are 67, 169, and 510 meV, respectively (see Table S7 for all the $E_a^{NEB}$). These activation energies from NEB can be directly related to *D* via the Arrhenius equation.

We first examined the impact of cation disorder on diffusion in the T[3×4] structure. Note that there is a detailed description of the barriers provided in the SI for the pristine, cation-disordered and block size variation models, as well as a table for the corresponding values between the relevant sites in Table S7. Here, we only provide a summary of the influence of the different structural types on the fastest diffusion sites. These disordered structures can compare how the placement of Mo influences $E_a^{NEB}$. We provide a comparison of *D* for these window sites which have the fastest diffusion. When Mo is placed at the center of the block T[3×4-dis-1], *D* becomes faster by two times in the "middle" windows, and slower by two orders of magnitude in the windows "near" Mo (Figure 5b and Table S10) as compared to pristine T[3×4]. In contrast, the windows "far" from the Mo have the diffusion of the same order of the magnitude as pristine T[3×4], but 1.5 times slower (Table S10). Furthermore, in the case of the SQS structure (T[3×4-dis-5]) where both Mo are at the corners of the blocks across the shear plane, diffusion remains of the same order of magnitude as pristine T[3×4] (Table S11). Overall, these trends indicate that the placement of Mo in the block slows down the fastest diffusion sites, whereas the placement at the shear plane octahedra does not influence the diffusion.

For the T[3×5] block, which contains an O-vacancy for charge-balancing as forementioned, we observed that the diffusion from the "near-middle" and "far-middle" windows were slower by an order of magnitude compared to pristine T[3×4], and the diffusion from other windows gets 1.3 times faster in the block without an O-vacancy (Figure 5c and Table S12). Thus, to test the impact of the oxygen vacancy alone, we find that in the T[3×5], the diffusion in middle windows that contain the O-vacancy, i.e., the windows near this O-vacancy (see Figure 5c) becomes slower by two orders of magnitude in comparison to pristine T[3×4] via an increased NEB barrier of 528 meV. In contrast, diffusion in windows far from the O-vacancy are similar, but 1.7 times slower. For the mixed phase T[3×4-3×5], the Li sites are a combination of sites of T[3×4] and T[3×5] blocks and have similar diffusion from the sites present in T[3×4] and T[3×5] blocks. Overall, these trends indicate that the O-vacancy slows down the fastest diffusion paths, whereas an increase in block size enhances the window site diffusion (exception: middle windows).

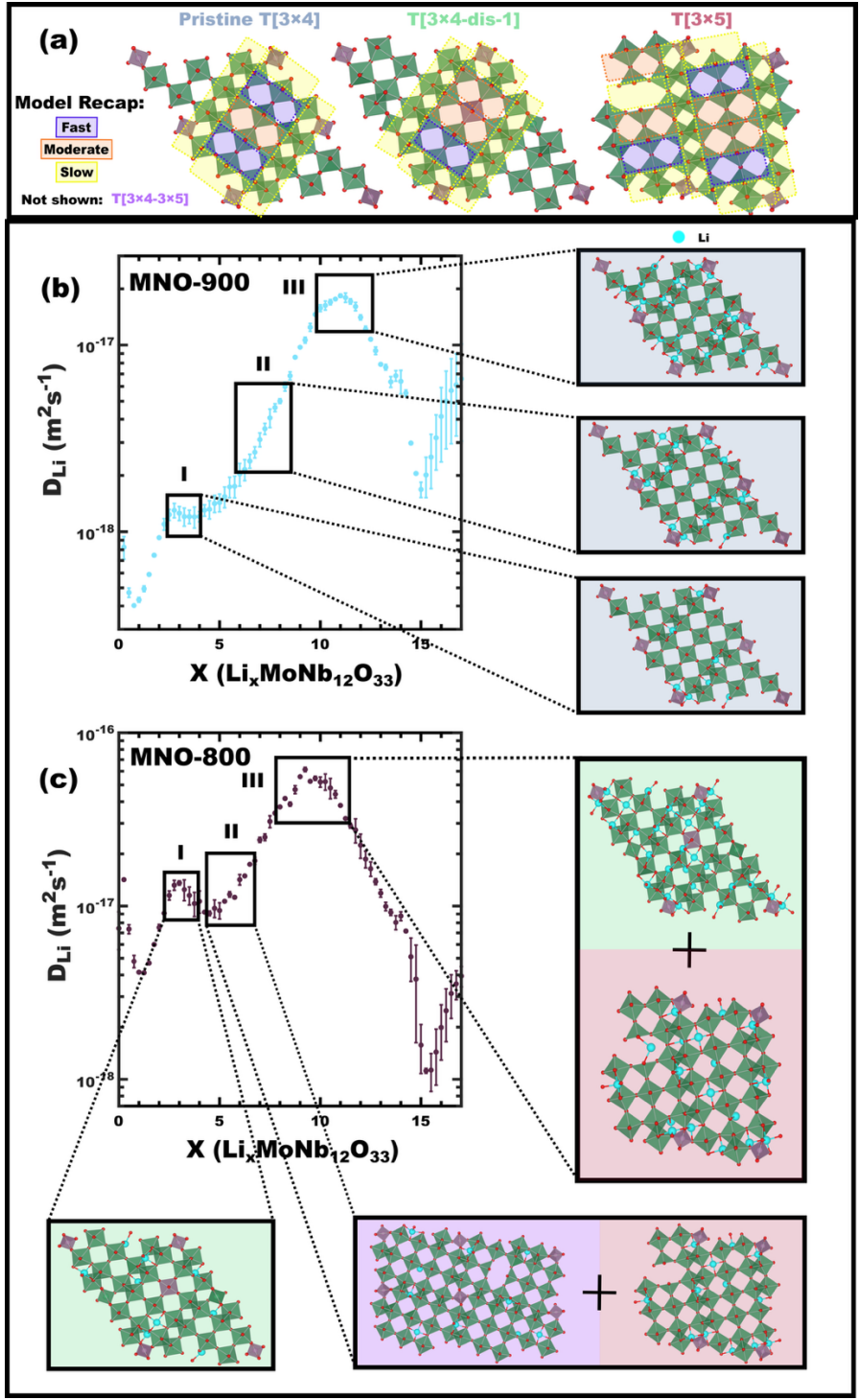

Figure 6. Computational transport results are compared to experimental results. The relative calculated diffusion rates were highlighted for specific Li sites (a). Ground state Li configurations were calculated for each model throughout lithiation (Figure S14). The experimental diffusion profile for MNO-900 (b) is compared to the pristine T[3×4] model where slow pocket diffusion sites are first occupied, followed by moderate speed middle windows of the block, and finally fast sites when other windows are occupied. Likewise, MNO-800 data (c) was compared to defect-rich models where faster sites were occupied starting from lower lithium concentrations.

Understanding lithium diffusion further requires consideration of the sequence of lithium site occupations for each model. As mentioned above, there are different sites that have substantially different individual $D$ values (Figure 6a), however these are only relevant when occupied with Li. The sequential filling of Li sites were identified using CE combined with metropolis Monte Carlo (MC) simulations at a higher temperature. The Li-fillings for the ground state of pristine T[3×4], T[3×4-dis-1], T[3×5], and T[3×4-3×5] is shown in Figure S14. In pristine T[3×4], the Ps and Pe pocket sites are first occupied at $x \leq 6.5$ (Figure 6b inset I) which notably have relatively slow diffusion (Figure 6a). While the Wv site in the middle of the block is occupied for $x \geq 1.5$, this site has relatively moderate diffusivity. When $x \geq 7.5$ (Figure 6b inset II) the Pc and Wh sites in the middle of the block start being occupied. It is not until $x \geq 11.5$ (Figure 6b inset III) and $x \geq 14.5$ that faster diffusion sites including the Wv and other Wh sites start becoming occupied, respectively. Thus, the lithiation of pristine T[3×4] progressively places lithium on sites with faster and faster diffusivity, similar to the experimental trend (Figure 6b). In contrast, the T[3×4-dis-1] model populates the fast diffusion sites Wh and Wv ("far" from Mo) at $x \geq 3.5$, and $x \geq 8.5$, respectively (Figure 6c inset I). Notably both of these fast sites are populated with lower Li concentrations than the pristine T[3×4] model. Similarly, both the T[3×5] and T[3×4-3×5] models start occupying the fastest Wh sites with $x \geq 6$ and $x \geq 5$, respectively (Figure 6c inset II). Curiously both models with cation disorder and block size variation (with fixed stoichiometry) led to the occupation of the fastest Wh and Wv sites at lower extents of lithiation than the pristine model.

Table 1. Diffusion coefficient (in $m^2 s^{-1}$) at different concentrations of Li for the considered structural models.

| Li Concentrations | Pristine T[3×4] | T[3×4-dis-1] | T[3×5] | T[3×4-3×5] |
|---|---|---|---|---|
| 25% | 2.00x10$^{-12}$ | 3.07x10$^{-10}$ | 3.08x10$^{-11}$ | 3.26x10$^{-11}$ |
| 50% | 3.12x10$^{-13}$ | 3.01x10$^{-11}$ | 1.30x10$^{-11}$ | 1.57x10$^{-11}$ |
| 75% | 3.17x10$^{-12}$ | 5.93x10$^{-13}$ | 2.40x10$^{-12}$ | 8.95x10$^{-12}$ |

The faster diffusion is confirmed with MD simulations performed at higher concentrations of lithiation, performed at 300 K. $D(x)$ was calculated at different concentrations of Li, specifically $x$ values corresponding to ~25, 50 and 75 at% Li per transition metal. To provide accurate estimates of the relative values for each Li concentration, we averaged $D$ starting from three lowest energy configurations found from CE approach at each concentration (Table 1). $D$ is higher in comparison to the pristine T[3×4] at 25 and 50% concentrations, in the order T[3×4-dis-1] > T[3×4-3×5] > T[3×5] > T[3×4]. Whereas at a higher concentration of 75 at% Li, $D$ in all structures remains roughly comparable with T[3×4]. This trend is notably consistent with several experimental observations, namely the ~1 order of magnitude faster diffusivity for MNO-800 (Figure 6c insets I-III) for early stages of lithiation (x<10) and then the convergence with MNO-900 diffusivity values at late stages of lithiation (x>12). The wider second parabola (peak at x~3) for MNO-800

(Figure 4e, shaded regions) is also consistent with additional Wh/Wv lithium sites being populated at lower extents of lithiation. Furthermore, the subsequently narrower third parabola (peak at x~10) is consistent with partial occupation and thus fewer unoccupied Wh/Wv sites remaining. Thus, the overall diffusivity trends are understandable from the perspective of modest changes to site-diffusivities with substantial changes to the sequence of Li site occupation.

**Conclusion**
Wadsley-Roth (WR) niobates have shown remarkable capability as fast-charging anode materials that combine rapid ionic diffusion with good electronic conductivity and potential for long-term cycling. WR compounds are known to sometimes be enhanced by defects from limited annealing, however such materials often contain both Wadsley defects (variable block size) and transition metal disorder where the corresponding effects have not yet been parsed or understood mechanistically. Here $MoNb_{12}O_{33}$ was prepared at two different temperatures selected to compare a defect-rich condition (MNO-800) with a proximal order-rich condition (MNO-900) as assessed by thorough XRD/XANES/EXAFS and STEM characterizations. Lithium half cells were galvanostatically cycled where MNO-800 exhibited enhanced capacity (307 mAh/g at C/10, 4.66% higher) and significantly improved rate capability with 200 mAh/g at 10C. Additionally, the defect-rich MNO-800 possessed remarkable cycling stability with ~93% and ~74% capacity retention thru 1k and 10k cycles at 20C, respectively. ICI overpotential analysis identified solid state diffusion as the rate limiting process where MNO-800 correspondingly exhibited ~3X faster capacity-weighted diffusivity. DFT and MLIP analysis were used to examine the possible roles of Wadsley defects and transition metal disorder where both defect model-types led to fast diffusion sites becoming occupied and activated at lower extents of lithiation. Therefore, both defect types were attributed to the enhanced performance in MNO-800 and offering a promising strategy to accelerate diffusivity in other intercalation materials.

**Experimental**
*Chemicals*
Niobium ethoxide (NbOEth, Sigma-Aldrich, 99.5%), Molybdenum (VI) dioxide bis(acetylacetonate) (Bis-Moly, Strem, 95% min), Nitric acid (Sigma-Aldrich, 70%), Ethanol (Decon Labs, 100%), Super P Conductive Carbon Black ("carbon black," MSE), N-Methyl-2-pyrrolidone (NMP, Sigma-Aldrich, 99.5%), Poly-vinylidene fluoride (PVDF, MTI, 99.5%), Acetone (Fisher, 99.5%), Lithium hexafluorophosphate in 1:1 ethylene carbonate/dimethyl carbonate (Sigma-Aldrich), Lithium metal (MSE, 99.9%) were used as received. The oxide precursors were stored in an Ar gas filled glovebox and transferred to a fume hood via air-free syringe for usage in atmosphere. Carbon-coated copper TEM grids (Electron Microscopy Sciences) were used as received. Kapton tape was used as a substrate for SAXS (McMaster-Carr). Tissue-Tek (Sakura) was used to mount samples for Cryo-ultramicrotomy (Leica, EM UC6).

*Synthesis*
$MoNb_{12}O_{33}$ was synthesized via an acid-stabilized sol-gel route. Niobium ethoxide and molybdenum dioxide bis(acetylacetonate) were combined in ethanol with a 12:1:20 molar ratio, respectively. Then, nitric acid (5:1 molar ratio of metal precursors to nitric acid) was added while the solution was stirred gently. Following 20 mins of moderate stirring, drops of water were slowly added until a solid gel formed. The gel was then heated at 105°C on a hot plate in a glass vial until a dry powder remained. The powder was ground in a mortar and pestle and separated into two aliquots for heat treatments in a furnace at 800 °C (MNO-800) and 900 °C (MNO-900). This crystallization treatment used a 5 °C/min ramp rate and was followed by a hold for 6 h followed by natural cooling.

*XRD/Rietveld Refinements*
Powder X-ray diffraction (XRD) was employed, with samples loaded into 0.3 mm diameter Kapton capillaries and sealed on both ends. The diffraction patterns were collected over the 5-60 °2θ range using a Bruker D8 Advance (Madison, WI, USA) instrument equipped with a Mo Kα radiation source (λ = 0.7093 Å) with convergent beam, incident beam monochromator, LynxEye detector with thick Si crystal, and rotating capillary sample stage. The crystal phase analysis was performed using MDI Jade (v.9.1) software, whereas the Rietveld refinement was done using Bruker Topas (v.7) to obtain information on the lattice parameters, volume, and crystallite size and microstrain. Both samples were matched to the same crystal phase available in the ICDD database (04-021-7212 $MoNb_{12}O_{33}$, monoclinic, space group $C2(5)$).[55] The angle regions around 22.5 and 31.8° 2θ were excluded from the Rietveld refinement because scattering from the beam stop used in the instrument setup caused overlapping peaks in the experimental data. Atom positions ($x$, $y$, $z$), site occupancy, and isotropic displacement parameters were not refined during the refinement procedure.

*EXAFS/XANES*
X-ray absorption spectroscopy (XAS) was used to gain an insight into the electronic structure and local structural features of absorber atoms. X-ray near-edge structure (XANES) and Extended X-ray absorption fine structure (EXAFS) spectra were collected at the 10-ID-B synchrotron beamline at the Advanced Photon Source at the Argonne National Laboratory.[56] The samples for ambient measurements were prepared by diluting MNO-800 or MNO-900 powder samples with an appropriate amount of BN dilutant powder and loading into nylon washers 19 mm in diameter. The samples were contained by Kapton tape on both sides of the washer. The Nb K edge (18,986 eV) and Mo K (20,000 eV) edge measurements were accompanied by Nb and Mo foils measurements, respectively, that were collected synchronously as references.

The collected spectra were analyzed using the Demeter software package (v.0.9.2.6).[57] The pre-processing of the collected data, including calibration, reference alignment, normalization, and background removal using $k^2$-weighted spline function, was performed in software Athena of the Demeter package running on the IFEFFIT algorithm. Each ambient spectrum was an average of three replicate measurements. The Nb K and Mo K edge positions ($E_0$) were determined from the position of the zero crossing of the second derivative and white line, respectively. Integrated AUTOBK algorithm was applied to minimize the effect of the background below 1 Å for Nb K edge and 0.8 Å for Mo K edge.

The pre-processed EXAFS oscillations $\chi(k)$ were presented as a function of photoelectron wavenumber $k$ ($k$-space) and then Fourier transformed (FT) into a radial distribution function, $\chi(R)$ ($R$-space). The Mo K edge EXAFS data were modeled against the theoretical $MoNb_{12}O_{33}$ structure (ICDD entry #04-021-7212) using software Artemis of the Demeter package, with a built-in FEFF6 algorithm. The modeling of the first coordination shell of MNO-900 utilized $k$, $k^2$, and $k^3$ weighing over a 0.8−3 Å R-range and 2.5−9.0 Å$^{-1}$ $k$-range. The modeling included fitting of the degeneracy (i.e., coordination number) of the scattering paths ($N$), energy shift ($\Delta E_0$), adjustment of the half-path length ($\Delta R$), and mean square displacement ($\sigma^2$). The amplitude reduction factor ($S_0^2$) remained fixed at 0.97, which was the value that was obtained for the Mo reference foil.

*Scanning Transmission Electron Microscopy*
A probe-corrected Titan operated at 200kV was used to acquire high annular dark-field (HAADF) STEM images with a 50 pA probe current and a convergence angle of 20mrad. Additionally, a probe-corrected JEOL NeoARM operated at 200kV was used to acquire STEM images with a 27 mrad convergence semi-angle, and a HAADF collection angle of 68 to 280 mrad. The EDS data was collected using the JEOL dual large solid angle detector. Electron-transparent samples were prepared via cryo-ultramicrotomy (Leica, EM UC6) with a nominal 50 nm thickness. Radial distribution analysis was completed using custom-made MATLAB code to select the atomic coordinates in each HAADF image. The resulting coordinates were analyzed using CORDERLY software from Guldin et al. to determine the radial distribution functions (RDF).[58]

*Scanning Electron Microscopy – Energy Dispersive Spectroscopy*
Energy Dispersive Spectroscopy (EDS) was performed using a Zeiss Gemini500 with an accelerating voltage of 30 keV and an EDAX Octane Elect Super detector. Quantification analysis of metal-metal ratios was completed by EDAX APEX software.

*Electrochemistry*
Electrodes were prepared from slurries for lithium half-cell measurements. Each sample's slurry was prepared by grinding an 8:1 mass ratio of active material to carbon black in a mortar and pestle. The ground carbon/oxide mixture was added to a 25 g/ml solution of PVDF in NMP to create an overall ratio of 8:1:1 of oxide, carbon black, and PVDF. Next, the NMP-based mixture was stirred for an hour to ensure thorough mixing. Copper foil current collectors were cleaned by wiping using Kimwipes moistened with acetone. The slurry was then doctor bladed onto the copper foils using a blade height of 15 μm. The freshly bladed films were immediately dried on a pre-heated hot plate set to 85 °C for 15 min. The films were further dried in a vacuum oven set to 100 °C for 22 hrs. After drying, the electrodes were punched into 12 mm discs, weighed in triplicate, and brought into an argon glovebox for assembly into coin cells. Coin cells were assembled using a 2032-type cases/lids, 1.0 M LiPF$_6$ in ethylene carbonate/dimethyl carbonate electrolyte, Whatman glass fiber separators, 16 mm lithium metal chips, two 0.5 mm stainless steel spacers, and 1 stainless steel wave spring. A Biologic BCS-810 was used for galvanostatic cycling and intermittent current interruption (ICI).[44] The C-rate was defined based on one electron per transition metal, corresponding to a 1C current density of 200.4 mA/g for $MoNb_{12}O_{33}$. Galvanostatic cycling was conducted between 1.0-3.0 V vs Li/Li$^+$ with a voltage hold period between each cycle. After galvanostatic cycling, the ICI technique was used to calculate diffusivity values with a current density corresponding to 0.1C.[5] During ICI, each 300 s galvanostatic period was interrupted by changing the current to zero for 10 s. The corresponding voltage transient was monitored to derive the lithium diffusivity. Detailed overpotential measurements utilized the Bio-Logic modulobat technique to capture voltage transients with minimal delay. The voltage relaxation in the first 2 ms ($\Delta V_{2ms}$) was acquired. The charge transfer overpotential ($\Delta V_{CT}$) was determined as the difference between $V_{2ms}$ and the t=0 extrapolated diffusion potential (Figure S12).[45] Diffusion overpotential ($\Delta V_D$) was calculated by comparison to a pseudoequilibrium 0.05C GCD measurement where the minor hysteresis was accounted for by averaging the lithiation and delithiation voltages at each Li$_x$ extent.[46,59] Each condition was measured in triplicate to present average battery values along with the error-of-the-mean. The specific surface areas used in diffusivity calculations were derived from SAXS Porod analysis. Datasets published elsewhere were extracted with PlotDigitizer for inclusion in the Ragone comparisons (Table S4). Tap density was measured using a custom-made apparatus with machined steel cylinders following ISO standards for tapping requirements.[60]

*Small-Angle X-ray Scattering*
Small-angle X-ray scattering (SAXS) measurements were used for Porod surface area analysis on both samples. Measurements were performed on a SAXSLab Ganesha at the South Carolina SAXS Collaborative (SCSC). A Xenocs GeniX 3D microfocus source with a copper target to generate a monochromatic X-ray beam with a wavelength of 0.154 nm and an intensity of 5.54E7 counts per second was used. A Pilatus detector captured both the scattered signal and was used to measure the direct/transmitted beam intensities. Samples were mounted using Kapton tape and all data had the Kapton background signal subtracted using SAXSGUI software. The scattering data is presented where the magnitude of the momentum transfer vector q corresponds to q=4$\pi$sin($\theta$)/$\lambda$. Standardless absolute scattering intensities were reported with the sample thickness determined using the "apparent thickness" approach developed by Spalla et al.[47,61] The X-ray absorption coefficient of 468 cm$^{-1}$ (NIST calculator) was used with the measured transmissivities to calculate the apparent sample thicknesses.[62] Importantly, the X-ray collimation was kept constant during the transmission measurement so that the illuminated region was the same as that used during the SAXS measurement (powders have heterogeneous thickness). The counts/second for each pixel of the Pilatus detector were validated to be within the linear regime for quantitative accuracy. Plots were used to calculate the specific surface area starting from the surface to volume ratio defined by:

$$\Sigma = \frac{(I_{abs}q^4)}{2\pi(\Delta SLD)^2}$$

Where $\Delta$SLD was equal to the scattering length density of the material. The SLD value of 3.55E-5 A$^{-2}$ for MoNb$_{12}$O$_{33}$ was calculated using a NIST resource for the utilized X-ray wavelength used.[63] The measurable Porod region was the indicated on the plot as the constant $I_{abs}q^4$ portion of the scattering data where the high-q limit was constrained to values with less than 1% intensity error. The sample specific surface area (S) was then calculated using:

$$S = \frac{\Sigma}{\rho}$$

where **$\rho$** is the sample bulk density (4.55 g/cm$^3$).

*Operando Wide-Angle X-ray Scattering*
Operando wide-angle X-ray scattering (WAXS) was performed on the same SAXSLab Ganesha at the South Carolina SAXS Collaborative (SCSC) with the same specs as mentioned previously. Continuous measurements were broken into 30 min intervals as the cells were cycled galvanostatically at C/10 on a Biologic SP-150 potentiostat. Electrodes (as-made previously) were assembled vs lithium into custom-made 2032 type coin cells with 5mm holes sealed by Kapton.

**Computational**

*Density Functional Theory (DFT) Details*

Spin-polarized DFT calculations were carried out using the Vienna Ab initio simulation package (VASP).[64,65] The ion-electron interactions were described using projector-augmented wave (PAW) pseudopotentials.[65,66] Specifically, Li_sv, Nb_pv, Mo_pv and O PAW pseudopotentials were considered for Li, Nb, Mo, and O, respectively. A generalized gradient approximation (GGA) based Perdew-Burke-Ernzerhof (PBE) exchange-correlation functional with Hubbard parameter $U$ of 4.38 for Mo was chosen from the MPRelaxSet, which tabulates $U$ parameters calibrated using the approach described by Wang et al.[67–69] The plane-wave kinetic energy cutoff of 520 eV was used for basis functions. A Γ-centered k-point mesh with a grid density of at least 1000/(atoms/unit cell) was used for Brillouin zone integration. The geometries were optimized until the forces on each atom were smaller than 0.01 eV/Å and the energy converged within the threshold of 0.001 meV.

*Nudged Elastic Band (NEB)*

NEB calculations were performed in the cell containing two formula units.[70,71] A single Li atom was inserted at each unique position. In the case of input structures for NEB calculations, only the atomic positions were optimized to keep the cell fixed during the transition state search. In total, 7 images were used to resolve the path between different hops. The NEB calculations were stopped once the force normal to the reaction pathway on each image was smaller than 0.05 eV/Å.

*Machine-Learned Interatomic Potential (MLIP)*

The MLIP model is trained using the MACE architecture, an equivariant message-passing tensor network that combines a message-passing mechanism with high body order interactions.[72,73] We used a multi-head approach to fine-tune from the MACE-MP-0b3 model with reference energies chosen from the atomic energies.[74] In this approach, the model training resumes from the checkpoint of foundation model trained to the Materials Project, but have two heads for predictions, one for randomly selected 20,000 samples from Materials Project, and the other for the dataset created in this study.[68] For the construction of the training set for the MLIP, we first sampled 3165 configurations from DFT-relaxation trajectories that were also used for building the DFT-Cluster Expansion (CE) model. This MLIP was then used to run NEB calculations for Li hopping in pristine T[3×4] and T[3×5] structures (Table S8). To further improve the MLIP predictions, the training set was augmented with 712 samples taken from contour exploring performed at the MLIP transition states.[75] This method ensures the system remains at the high energy state but probe potential energy surface perpendicular to the reaction pathway. The retrained MLIP was then used for all the reported MD simulations, NEB calculations and to train an CE model for determining the Li ordering in this work. The CE model of the MLIP is used because it allows for a dramatic decrease in the computational cost for the determination of the Li ordering at various $Li_x$.

*Cluster Expansion*

To determine the ground state in different structures considered in this study at various Li concentrations, first we built cluster expansion (CE) models of the formation energy $E_{f,i}$ which is calculated using the following equation:

$$E_{f,i} = (E_i - E_{\text{host}} - x_{\text{Li}}E_{\text{Li}})/N$$

where $E_i$ is the energy of host cell containing Li atoms, $E_{host}$ is the energy of host cell, and $E_{Li}$ is the energy of the bulk Li.[76] $x_{Li}$ is the number of Li atoms and $N$ is the total number of atoms in the host cell. We used the Python package CELL to build the CE models. The details of CE models are provided in Table S13. Then, using these CE models, we enumerate structures from the Monte Carlo run at a high temperature of 2000 K with 100k steps. Subsequently, we optimized the lowest 1k structures with trained MLIP and found the lowest-energy state.

*Molecular dynamics*
MLIP-MD simulations were performed in the cell containing two formula units and Li is placed at different unique sites for 2 ns at 300 K. The simulations used a Langevin thermostat in the *NVT* ensemble, with a time step of 1 fs. The first 20 ps was kept for equilibration. From the rest of the trajectory, dividing it into subsamples of length 10 ps, the diffusion coefficient (*D*) was calculated from the Einstein relation:

$$D = \frac{1}{2d} \lim_{t \to \infty} \frac{d}{dt} \frac{1}{N} \sum_{i=1}^{N} \langle |r_i(t+t_0) - r_i(t_0)|^2 \rangle_{t_0}$$

where $N$ is the number of diffusing particles, $d$ is the dimensionality of the diffusion, and $r_i(t+t_0)$ represents the position of atom $i$ at time $t$, following an equilibration time $t_0$. Note that this equation is valid for so-called self-diffusion, which measures how a particle (e.g., Li ion) diffuses through a material in the absence of concentration gradients. Furthermore, MLIP-MD simulations were performed across five different temperatures in increments of 100 K (500–900 K), each for 2 ns for three lowest configurations at about 25%, 50%, and 75% Li per transition metal for all the structural models. The average room temperature (*T* = 300 K) *D* values were extracted by extrapolating the Arrhenius equation, which describes how the diffusion rate depends on temperature *T*:

$$D = D_0 e^{-\frac{E_a}{k_B T}}$$

where $E_a$ is the activation energy, and $k_B$ is the Boltzmann constant.

**Acknowledgements**
CS, MK, NK, IM, SM, CS, and MS acknowledge DOE support (DE-SC0023377). This work made use of the South Carolina SAXS Collaborative (SCSC). MRCAT operations are supported by the Department of Energy and the MRCAT member institutions. This research used resources of the Advanced Photon Source, a U.S. Department of Energy (DOE) Office of Science User Facility operated for the DOE Office of Science by Argonne National Laboratory under Contract No. DE-AC02-06CH11357. X-ray Photoelectron Spectroscopy (XPS) data were collected at the University of South Carolina XPS Facility (RRID: SCR_026176), financially supported by the Office of the Vice President for Research at the University of South Carolina. STEM work was performed in part at the Analytical Instrumentation Facility (AIF) at North Carolina State University, which is supported by the State of North Carolina and the National Science Foundation (award number ECCS-2025064). The AIF is a member of the North Carolina Research Triangle Nanotechnology Network (RTNN), a site in the National Nanotechnology Coordinated Infrastructure (NNCI). STEM work was also conducted as part of a user project was supported by the Center for Nanophase Materials Sciences (CNMS), which is a US Department of Energy, Office of Science User Facility at Oak Ridge National Laboratory. This research was conducted using instrumentation within ORNL's Materials Characterization Core provided by UT-Battelle, LLC under Contract No. DE-AC05-00OR22725 with the U.S. Department of Energy.

**References**
(1) Yang, Y.; Zhao, J. Wadsley–Roth Crystallographic Shear Structure Niobium-Based Oxides: Promising Anode Materials for High-Safety Lithium-Ion Batteries. *Adv. Sci.* **2021**, *8* (12), 2004855. https://doi.org/10.1002/advs.202004855.
(2) Griffith, K. J.; Wiaderek, K. M.; Cibin, G.; Marbella, L. E.; Grey, C. P. Niobium Tungsten Oxides for High-Rate Lithium-Ion Energy Storage. *Nature* **2018**, *559* (7715), 556–563. https://doi.org/10.1038/s41586-018-0347-0.
(3) Saber, M.; Reynolds, C.; Li, J.; Pollock, T. M.; Van der Ven, A. Chemical and Structural Factors Affecting the Stability of Wadsley–Roth Block Phases. *Inorg. Chem.* **2023**, *62* (42), 17317–17332. https://doi.org/10.1021/acs.inorgchem.3c02595.
(4) Cava, R. J.; Murphy, D. W.; Zahurak, S. M. Lithium Insertion in Wadsley-Roth Phases Based on Niobium Oxide. *J. Electrochem. Soc.* **1983**, *130* (12), 2345–2351. https://doi.org/10.1149/1.2119583.
(5) Griffith, K. J.; Seymour, I. D.; Hope, M. A.; Butala, M. M.; Lamontagne, L. K.; Preefer, M. B.; Koçer, C. P.; Henkelman, G.; Morris, A. J.; Cliffe, M. J.; Dutton, S. E.; Grey, C. P. Ionic and Electronic Conduction in TiNb2O7. *J. Am. Chem. Soc.* **2019**, *141* (42), 16706–16725. https://doi.org/10.1021/jacs.9b06669.
(6) Muhit, M. A. A.; Wechsler, S. C.; Bare, Z. J. L.; Sturgill, C.; Keerthisinghe, N.; Grasser, M. A.; Morrison, G.; Sutton, C.; Stefik, M.; zur Loye, H.-C. Comparison of Lithium Diffusion in Isostructural Ta12MoO33 and Nb12MoO33: Experimental and Computational Insights from Single Crystals. *Chem. Mater.* **2024**, *36* (21), 10626–10639. https://doi.org/10.1021/acs.chemmater.4c02118.
(7) Koçer, C. P.; Griffith, K. J.; Grey, C. P.; Morris, A. J. Lithium Diffusion in Niobium Tungsten Oxide Shear Structures. *Chem. Mater.* **2020**, *32* (9), 3980–3989. https://doi.org/10.1021/acs.chemmater.0c00483.

(8) Zuras, E. J.; Fauth, F.; Rousse, G.; Grimaud, A. Investigating Lithium Ordering and Electronic Evolutions in Wadsley–Roth Phases. *J. Phys. Chem. C* **2025**, *129* (7), 3446–3456. https://doi.org/10.1021/acs.jpcc.4c08417.

(9) Koçer, C. P.; Griffith, K. J.; Grey, C. P.; Morris, A. J. Cation Disorder and Lithium Insertion Mechanism of Wadsley–Roth Crystallographic Shear Phases from First Principles. *J. Am. Chem. Soc.* **2019**, *141* (38), 15121–15134. https://doi.org/10.1021/jacs.9b06316.

(10) Zhu, X.; Xu, J.; Luo, Y.; Fu, Q.; Liang, G.; Luo, L.; Chen, Y.; Lin, C.; Zhao, X. S. MoNb12O33 as a New Anode Material for High-Capacity, Safe, Rapid and Durable Li+ Storage: Structural Characteristics, Electrochemical Properties and Working Mechanisms. *J. Mater. Chem. A* **2019**, *7* (11), 6522–6532. https://doi.org/10.1039/C9TA00309F.

(11) Reynaud, M.; Serrano-Sevillano, J.; Casas-Cabanas, M. Imperfect Battery Materials: A Closer Look at the Role of Defects in Electrochemical Performance. *Chem. Mater.* **2023**, *35* (9), 3345–3363. https://doi.org/10.1021/acs.chemmater.2c03481.

(12) Maier, J. Review—Battery Materials: Why Defect Chemistry? *J. Electrochem. Soc.* **2015**, *162* (14), A2380–A2386. https://doi.org/10.1149/2.0011514jes.

(13) Sturgill, C.; Milisavljevic, I.; Wechsler, S. C.; Muhit, M. A. A.; zur Loye, H.-C.; Misture, S.; Stefik, M. Tailored TiNb2O7 Particle Size, Defects, and Crystallinity Accelerate Lithiation. *Chem. Mater.* **2025**, *37* (2), 624–635. https://doi.org/10.1021/acs.chemmater.4c02279.

(14) van den Bergh, W.; Wechsler, S.; Lokupitiya, H. N.; Jarocha, L.; Kim, K.; Chapman, J.; Kweon, K. E.; Wood, B. C.; Heald, S.; Stefik, M. Amorphization of Pseudocapacitive T−Nb2O5 Accelerates Lithium Diffusivity as Revealed Using Tunable Isomorphic Architectures. *Batter. Supercaps* **2022**, *5* (6), e202200056. https://doi.org/10.1002/batt.202200056.

(15) Wu, Q.; Kang, Y.; Chen, G.; Chen, J.; Chen, M.; Li, W.; Lv, Z.; Yang, H.; Lin, P.; Qiao, Y.; Zhao, J.; Yang, Y. Ultrafast Carbothermal Shock Synthesis of Wadsley–Roth Phase Niobium-Based Oxides for Fast-Charging Lithium-Ion Batteries. *Adv. Funct. Mater.* **2024**, *34* (23), 2315248. https://doi.org/10.1002/adfm.202315248.

(16) Allpress, J. G. Mixed Oxides of Titanium and Niobium: Defects in Quenched Samples. *J. Solid State Chem.* **1970**, *2* (1), 78–93. https://doi.org/10.1016/0022-4596(70)90037-X.

(17) Allpress, J. G.; Roth, R. S. The Effect of Annealing on the Concentration of Wadsley Defects in the Nb2O5 WO3 System. *J. Solid State Chem.* **1971**, *3* (2), 209–216. https://doi.org/10.1016/0022-4596(71)90030-2.

(18) Cao, J.; Xia, J.; Li, X.; Li, Y.; Liu, P.; Tian, L.; Qiao, P.; Liu, C.; Wang, Y.; Meng, X. Defect-Mediated Growth of Crystallographic Shear Plane. *Small* **2023**, *19* (44), 2302365. https://doi.org/10.1002/smll.202302365.

(19) Andersson, S.; Galy, J. Wadsley Defects and Crystallographic Shear in Hexagonally Close-Packed Structures. *J. Solid State Chem.* **1970**, *1* (3–4), 576–582. https://doi.org/10.1016/0022-4596(70)90144-1.

(20) Eyring, L. Structure, Defects, and Nonstoichiometry in Oxides: An Electron Microscopic View*. In *Nonstoichiometric Oxides*; Sørensen, O. T., Ed.; Academic Press, 1981; pp 337–398. https://doi.org/10.1016/B978-0-12-655280-5.50012-2.

(21) Voskanyan, A. A.; Abramchuk, M.; Navrotsky, A. Entropy Stabilization of TiO2–Nb2O5 Wadsley–Roth Shear Phases and Their Prospects for Lithium-Ion Battery Anode Materials. *Chem. Mater.* **2020**, *32* (12), 5301–5308. https://doi.org/10.1021/acs.chemmater.0c01553.

(22) Ahn, Y.; Li, T.; Huang, S.; Ding, Y.; Hwang, S.; Wang, W.; Luo, Z.; Wang, J.; Nam, G.; Liu, M. Mixed-Phase Niobium Oxide as a Durable and Ultra-Fast Charging Anode for High-Power Lithium-Ion Batteries. *Adv. Funct. Mater.* **2024**, *34* (8), 2310853. https://doi.org/10.1002/adfm.202310853.

(23) Li, T.; Nam, G.; Liu, K.; Wang, J.-H.; Zhao, B.; Ding, Y.; Soule, L.; Avdeev, M.; Luo, Z.; Zhang, W.; Yuan, T.; Jing, P.; Kim, M. G.; Song, Y.; Liu, M. A Niobium Oxide with a Shear Structure and Planar Defects for High-Power Lithium Ion Batteries. *Energy Environ. Sci.* **2022**, *15* (1), 254–264. https://doi.org/10.1039/D1EE02664J.

(24) Salzer, L. D.; Diamond, B.; Nieto, K.; Evans, R. C.; Prieto, A. L.; Sambur, J. B. Structure–Property Relationships in High-Rate Anode Materials Based on Niobium Tungsten Oxide Shear Structures. *ACS Appl. Energy Mater.* **2023**, *6* (3), 1685–1691. https://doi.org/10.1021/acsaem.2c03573.

(25) Jing, P.; Liu, M.; Ho, H.-P.; Ma, Y.; Hua, W.; Li, H.; Guo, N.; Ding, Y.; Zhang, W.; Chen, H.; Zhao, B.; Wang, J.; Liu, M. Tailoring the Wadsley–Roth Crystallographic Shear Structures for High-Power Lithium-Ion Batteries. *Energy Environ. Sci.* **2024**, *17* (18), 6571–6581. https://doi.org/10.1039/D4EE02293A.

(26) Cheetham, A. K.; Von Dreele, R. B. Cation Distributions in Niobium Oxide Block Structures. *Nat. Phys. Sci.* **1973**, *244* (139), 139–140. https://doi.org/10.1038/physci244139a0.

(27) Ma, J.; Xiang, Y.; Xu, J.; Zhang, W.; Zhang, H.; Qiu, J.; Zhu, X.; Zhang, H.; Lin, H.; Cao, G. Reducing Lithium-Diffusion Barrier on the Wadsley–Roth Crystallographic Shear Plane via Low-Valent Cation Doping for Ultrahigh Power Lithium-Ion Batteries. *Adv. Energy Mater.* **2025**, *15* (12), 2403623. https://doi.org/10.1002/aenm.202403623.

(28) Ahn, Y.; Hu, X.; Ding, Y.; Kim, C.; Wu, Y.-C.; Kim, T.; Liu, M. Synergistic Effects of Entropy Tuning in Niobium-Based Oxide Anode for Fast-Charging Lithium-Ion Batteries. *Adv. Funct. Mater.* n/a (n/a), e09533. https://doi.org/10.1002/adfm.202509533.

(29) Abdellahi, A.; Urban, A.; Dacek, S.; Ceder, G. Understanding the Effect of Cation Disorder on the Voltage Profile of Lithium Transition-Metal Oxides. *Chem. Mater.* **2016**, *28* (15), 5373–5383. https://doi.org/10.1021/acs.chemmater.6b01438.

(30) Ji, H.; Urban, A.; Kitchaev, D. A.; Kwon, D.-H.; Artrith, N.; Ophus, C.; Huang, W.; Cai, Z.; Shi, T.; Kim, J. C.; Kim, H.; Ceder, G. Hidden Structural and Chemical Order Controls Lithium Transport in Cation-Disordered Oxides for Rechargeable Batteries. *Nat. Commun.* **2019**, *10* (1), 592. https://doi.org/10.1038/s41467-019-08490-w.

(31) Driscoll, E. H.; Green, A.; Fortes, D.; Howard, C.; Driscoll, L. L.; Kendrick, E.; Greaves, C.; Slater, P. R. A High-Power 4 × 4: Crystallographic and Electrochemical Insights into a Novel Wadsley–Roth Anode Nb9Ti1.5W1.5O30. *Chem. Commun.* **2024**, *60* (73), 10001–10004. https://doi.org/10.1039/D4CC02844A.

(32) Green, A. J.; Driscoll, E. H.; Lakhdar, Y.; Kendrick, E.; Slater, P. R. Structural and Electrochemical Insights into Novel Wadsley Roth Nb7Ti1.5Mo1.5O25 and Ta7Ti1.5Mo1.5O25 Anodes for Li-Ion Battery Application. *Dalton Trans.* **2023**, *52* (37), 13110–13118. https://doi.org/10.1039/D3DT02144K.

(33) Liu, Y.; Guilherme Buzanich, A.; Montoro, L. A.; Liu, H.; Liu, Y.; Emmerling, F.; Russo, P. A.; Pinna, N. A Partially Disordered Crystallographic Shear Block Structure as Fast-Charging Negative Electrode Material for Lithium-Ion Batteries. *Nat. Commun.* **2025**, *16* (1), 6507. https://doi.org/10.1038/s41467-025-61646-9.

(34) Iijima, S. Direct Observation of Lattice Defects in $H$-Nb$_2$O$_5$ by High Resolution Electron Microscopy. *Acta Crystallogr. Sect. A* **1973**, *29* (1), 18–24. https://doi.org/10.1107/S0567739473000045.

(35) Anderson, J. S. On Infinitely Adaptive Structures. *J. Chem. Soc. Dalton Trans.* **1973**, No. 10, 1107–1115. https://doi.org/10.1039/DT9730001107.

(36) Li, T.; Nam, G.; Liu, K.; Wang, J.-H.; Zhao, B.; Ding, Y.; Soule, L.; Avdeev, M.; Luo, Z.; Zhang, W.; Yuan, T.; Jing, P.; Kim, M. G.; Song, Y.; Liu, M. A Niobium Oxide with a Shear Structure and Planar Defects for High-Power Lithium Ion Batteries. *Energy Environ. Sci.* **2022**, *15* (1), 254–264. https://doi.org/10.1039/D1EE02664J.

(37) Qian, S.; Yu, H.; Yan, L.; Zhu, H.; Cheng, X.; Xie, Y.; Long, N.; Shui, M.; Shu, J. High-Rate Long-Life Pored Nanoribbon VNb9O25 Built by Interconnected Ultrafine Nanoparticles as Anode for Lithium-Ion Batteries. *ACS Appl. Mater. Interfaces* **2017**, *9* (36), 30608–30616. https://doi.org/10.1021/acsami.7b07460.

(38) Merryweather, A. J.; Jacquet, Q.; Emge, S. P.; Schnedermann, C.; Rao, A.; Grey, C. P. Operando Monitoring of Single-Particle Kinetic State-of-Charge Heterogeneities and Cracking in High-Rate Li-Ion Anodes. *Nat. Mater.* **2022**, *21* (11), 1306–1313. https://doi.org/10.1038/s41563-022-01324-z.

(39) Griffith, K. J.; Grey, C. P. Superionic Lithium Intercalation through 2 × 2 Nm2 Columns in the Crystallographic Shear Phase Nb18W8O69. *Chem. Mater.* **2020**, *32* (9), 3860–3868. https://doi.org/10.1021/acs.chemmater.9b05403.

(40) Griffith, K. J.; Seymour, I. D.; Hope, M. A.; Butala, M. M.; Lamontagne, L. K.; Preefer, M. B.; Koçer, C. P.; Henkelman, G.; Morris, A. J.; Cliffe, M. J.; Dutton, S. E.; Grey, C. P. Ionic and Electronic Conduction in TiNb2O7. *J. Am. Chem. Soc.* **2019**, *141* (42), 16706–16725. https://doi.org/10.1021/jacs.9b06669.

(41) Griffith, K. J.; Forse, A. C.; Griffin, J. M.; Grey, C. P. High-Rate Intercalation without Nanostructuring in Metastable Nb2O5 Bronze Phases. *J. Am. Chem. Soc.* **2016**, *138* (28), 8888–8899. https://doi.org/10.1021/jacs.6b04345.

(42) Jung, H.-G.; Jang, M. W.; Hassoun, J.; Sun, Y.-K.; Scrosati, B. A High-Rate Long-Life Li4Ti5O12/Li[Ni0.45Co0.1Mn1.45]O4 Lithium-Ion Battery. *Nat. Commun.* **2011**, *2* (1), 516. https://doi.org/10.1038/ncomms1527.

(43) Xiang, W.; Chen, M.; Zhou, X.; Chen, J.; Huang, H.; Sun, Z.; Lu, Y.; Zhang, G.; Wen, X.; Li, W. Highly Enforced Rate Capability of a Graphite Anode via Interphase Chemistry Tailoring Based on an Electrolyte Additive. *J. Phys. Chem. Lett.* **2022**, *13* (23), 5151–5159. https://doi.org/10.1021/acs.jpclett.2c01183.

(44) Chien, Y.-C.; Liu, H.; Menon, A. S.; Brant, W. R.; Brandell, D.; Lacey, M. J. Rapid Determination of Solid-State Diffusion Coefficients in Li-Based Batteries via Intermittent Current Interruption Method. *Nat. Commun.* **2023**, *14* (1), 2289. https://doi.org/10.1038/s41467-023-37989-6.

(45) Geng, Z.; Thiringer, T.; Lacey, M. J. Intermittent Current Interruption Method for Commercial Lithium-Ion Batteries Aging Characterization. *IEEE Trans. Transp. Electrification* **2022**, *8* (2), 2985–2995. https://doi.org/10.1109/TTE.2021.3125418.

(46) Kumar, M.; Muhit, M. A. A.; Sturgill, C.; Karimitari, N.; Barber, J. T.; Tisdale, H.; Stefik, M.; zur Loye, H.-C.; Sutton, C. Combined Experimental and Computational Analysis of Lithium Diffusion in Isostructural Pair VNb9O25 and VTa9O25. *ACS Appl. Energy Mater.* **2025**, *8* (18), 13407–13420. https://doi.org/10.1021/acsaem.5c01738.

(47) Spalla, O.; Lyonnard, S.; Testard, F. Analysis of the Small-Angle Intensity Scattered by a Porous and Granular Medium. *J. Appl. Crystallogr.* **2003**, *36* (2), 338–347. https://doi.org/10.1107/S0021889803002279.

(48) Schlumberger, C.; Scherdel, C.; Kriesten, M.; Leicht, P.; Keilbach, A.; Ehmann, H.; Kotnik, P.; Reichenauer, G.; Thommes, M. Reliable Surface Area Determination of Powders and Meso/Macroporous Materials: Small-Angle X-Ray Scattering and Gas Physisorption. *Microporous Mesoporous Mater.* **2022**, *329*, 111554. https://doi.org/10.1016/j.micromeso.2021.111554.

(49) Conway, B. E. Two-Dimensional and Quasi-Two-Dimensional Isotherms for Li Intercalation and *Upd* Processes at Surfaces. *Electrochimica Acta* **1993**, *38* (9), 1249–1258. https://doi.org/10.1016/0013-4686(93)80055-5.

(50) Prosini, P. P.; Lisi, M.; Zane, D.; Pasquali, M. Determination of the Chemical Diffusion Coefficient of Lithium in LiFePO4. *Solid State Ion.* **2002**, *148* (1), 45–51. https://doi.org/10.1016/S0167-2738(02)00134-0.

(51) Sturgill, C. J.; Sutton, C.; Schwenzel, J.; Stefik, M. Capacity-Weighted Figures-of-Merit for Battery Transport Metrics. *J. Mater. Chem. A* **2024**. https://doi.org/10.1039/D4TA06041E.

(52) Zunger, A.; Wei, S.-H.; Ferreira, L. G.; Bernard, J. E. Special Quasirandom Structures. *Phys. Rev. Lett.* **1990**, *65* (3), 353–356. https://doi.org/10.1103/PhysRevLett.65.353.

(53) van de Walle, A.; Tiwary, P.; de Jong, M.; Olmsted, D. L.; Asta, M.; Dick, A.; Shin, D.; Wang, Y.; Chen, L.-Q.; Liu, Z.-K. Efficient Stochastic Generation of Special Quasirandom Structures. *Calphad* **2013**, *42*, 13–18. https://doi.org/10.1016/j.calphad.2013.06.006.

(54) Gatehouse, B. M.; Wadsley, A. D. The Crystal Structure of the High Temperature Form of Niobium Pentoxide. *Acta Crystallogr.* **1964**, *17* (12), 1545–1554. https://doi.org/10.1107/S0365110X6400384X.

(55) Tabero, P. Thermal Expansion of Phases Formed in the System Nb2O5-MoO3. *J. Therm. Anal. Calorim.* **2003**, *74* (2), 491–496. https://doi.org/10.1023/B:JTAN.0000005185.84636.cc.

(56) Segre, C. U.; Leyarovska, N. E.; Chapman, L. D.; Lavender, W. M.; Plag, P. W.; King, A. S.; Kropf, A. J.; Bunker, B. A.; Kemner, K. M.; Dutta, P.; Duran, R. S.; Kaduk, J. The MRCAT Insertion Device Beamline at the Advanced Photon Source. *AIP Conf. Proc.* **2000**, *521* (1), 419–422. https://doi.org/10.1063/1.1291825.

(57) Ravel, B.; Newville, M. ATHENA, ARTEMIS, HEPHAESTUS: Data Analysis for X-Ray Absorption Spectroscopy Using IFEFFIT. *J. Synchrotron Radiat.* **2005**, *12* (4), 537–541. https://doi.org/10.1107/S0909049505012719.

(58) Mac Fhionnlaoich, N.; Qi, R.; Guldin, S. Application of the Spatial Distribution Function to Colloidal Ordering. *Langmuir* **2019**, *35* (50), 16605–16611. https://doi.org/10.1021/acs.langmuir.9b02877.

(59) Dreyer, W.; Jamnik, J.; Guhlke, C.; Huth, R.; Moškon, J.; Gaberšček, M. The Thermodynamic Origin of Hysteresis in Insertion Batteries. *Nat. Mater.* **2010**, *9* (5), 448–453. https://doi.org/10.1038/nmat2730.

(60) Metallic Powders — Determination of Tap Density. *Int. Organ. Stand.* **2011**, *ISO 3953* (4), 4.

(61) Joensen, K. Ganesha Operating Manual. https://smif.pratt.duke.edu/capabilities/55527 (accessed 2025-06-17).

(62) *NIST X-Ray Form Factor, Atten. Scatt. Tables Form Page*. https://physics.nist.gov/PhysRefData/FFast/html/form.html (accessed 2025-06-17).

(63) *Scattering Length Density Calculator*. https://www.ncnr.nist.gov/resources/sldcalc.html (accessed 2025-06-17).
(64) Kresse, G.; Furthmüller, J. Efficiency of Ab-Initio Total Energy Calculations for Metals and Semiconductors Using a Plane-Wave Basis Set. *Comput. Mater. Sci.* **1996**, *6* (1), 15–50. https://doi.org/10.1016/0927-0256(96)00008-0.
(65) Kresse, G.; Joubert, D. From Ultrasoft Pseudopotentials to the Projector Augmented-Wave Method. *Phys. Rev. B* **1999**, *59* (3), 1758–1775. https://doi.org/10.1103/PhysRevB.59.1758.
(66) Blöchl, P. E. Projector Augmented-Wave Method. *Phys. Rev. B* **1994**, *50* (24), 17953–17979. https://doi.org/10.1103/PhysRevB.50.17953.
(67) Perdew, J. P.; Burke, K.; Ernzerhof, M. Generalized Gradient Approximation Made Simple. *Phys. Rev. Lett.* **1996**, *77* (18), 3865–3868. https://doi.org/10.1103/PhysRevLett.77.3865.
(68) Jain, A.; Ong, S. P.; Hautier, G.; Chen, W.; Richards, W. D.; Dacek, S.; Cholia, S.; Gunter, D.; Skinner, D.; Ceder, G.; Persson, K. A. Commentary: The Materials Project: A Materials Genome Approach to Accelerating Materials Innovation. *APL Mater.* **2013**, *1* (1), 011002. https://doi.org/10.1063/1.4812323.
(69) Wang, L.; Maxisch, T.; Ceder, G. Oxidation Energies of Transition Metal Oxides within the $\mathrm{GGA}+\mathrm{U}$ Framework. *Phys. Rev. B* **2006**, *73* (19), 195107. https://doi.org/10.1103/PhysRevB.73.195107.
(70) Henkelman, G.; Uberuaga, B. P.; Jónsson, H. A Climbing Image Nudged Elastic Band Method for Finding Saddle Points and Minimum Energy Paths. *J. Chem. Phys.* **2000**, *113* (22), 9901–9904. https://doi.org/10.1063/1.1329672.
(71) Henkelman, G.; Jónsson, H. Improved Tangent Estimate in the Nudged Elastic Band Method for Finding Minimum Energy Paths and Saddle Points. *J. Chem. Phys.* **2000**, *113* (22), 9978–9985. https://doi.org/10.1063/1.1323224.
(72) Batatia, I.; Kovacs, D. P.; Simm, G.; Ortner, C.; Csanyi, G. MACE: Higher Order Equivariant Message Passing Neural Networks for Fast and Accurate Force Fields. *Adv. Neural Inf. Process. Syst.* **2022**, *35*, 11423–11436.
(73) Kovács, D. P.; Batatia, I.; Arany, E. S.; Csányi, G. Evaluation of the MACE Force Field Architecture: From Medicinal Chemistry to Materials Science. *J. Chem. Phys.* **2023**, *159* (4), 044118. https://doi.org/10.1063/5.0155322.
(74) Batatia, I.; Benner, P.; Chiang, Y.; Elena, A. M.; Kovács, D. P.; Riebesell, J.; Advincula, X. R.; Asta, M.; Avaylon, M.; Baldwin, W. J.; Berger, F.; Bernstein, N.; Bhowmik, A.; Bigi, F.; Blau, S. M.; Cărare, V.; Ceriotti, M.; Chong, S.; Darby, J. P.; De, S.; Pia, F. D.; Deringer, V. L.; Elijošius, R.; El-Machachi, Z.; Falcioni, F.; Fako, E.; Ferrari, A. C.; Gardner, J. L. A.; Gawkowski, M. J.; Genreith-Schriever, A.; George, J.; Goodall, R. E. A.; Grandel, J.; Grey, C. P.; Grigorev, P.; Han, S.; Handley, W.; Heenen, H. H.; Hermansson, K.; Holm, C.; Ho, C. H.; Hofmann, S.; Jaafar, J.; Jakob, K. S.; Jung, H.; Kapil, V.; Kaplan, A. D.; Karimitari, N.; Kermode, J. R.; Kourtis, P.; Kroupa, N.; Kullgren, J.; Kuner, M. C.; Kuryla, D.; Liepuoniute, G.; Lin, C.; Margraf, J. T.; Magdău, I.-B.; Michaelides, A.; Moore, J. H.; Naik, A. A.; Niblett, S. P.; Norwood, S. W.; O'Neill, N.; Ortner, C.; Persson, K. A.; Reuter, K.; Rosen, A. S.; Rosset, L. A. M.; Schaaf, L. L.; Schran, C.; Shi, B. X.; Sivonxay, E.; Stenczel, T. K.; Svahn, V.; Sutton, C.; Swinburne, T. D.; Tilly, J.; Oord, C. van der; Vargas, S.; Varga-Umbrich, E.; Vegge, T.; Vondrák, M.; Wang, Y.; Witt, W. C.; Wolf, T.; Zills, F.; Csányi, G. A Foundation Model for Atomistic Materials Chemistry. arXiv September 4, 2025. https://doi.org/10.48550/arXiv.2401.00096.

(75) Waters, M. J.; Rondinelli, J. M. Energy Contour Exploration with Potentiostatic Kinematics. *J. Phys. Condens. Matter* **2021**, *33* (44), 445901. https://doi.org/10.1088/1361-648X/ac1af0.

(76) Sanchez, J. M.; Ducastelle, F.; Gratias, D. Generalized Cluster Description of Multicomponent Systems. *Phys. Stat. Mech. Its Appl.* **1984**, *128* (1), 334–350. https://doi.org/10.1016/0378-4371(84)90096-7.